\documentstyle[preprint,tighten,aps,epsfig]{revtex}
\begin{document}
\title{Constituent quark model study of light- and strange-baryon spectra}
\author{A. Valcarce$^{(1)}$, H. Garcilazo$^{(2)}$, J. Vijande$^{(1)}$}
\address{$(1)$ Grupo de F{\'\i}sica Nuclear and IUFFyM, \\
Universidad de Salamanca, E-37008 Salamanca, Spain}
\address{$(2)$ Escuela Superior de F\'\i sica y Matem\'aticas, \\
Instituto Polit\'ecnico Nacional, Edificio 9,
07738 M\'exico D.F., Mexico}
\maketitle

\begin{abstract}
We investigate the structure of the $SU(3)$ octet and decuplet baryons
employing a constituent quark model designed for the study of the 
baryon-baryon interaction and successfully applied to the meson spectra. 
The model considers through the interacting potential perturbative, one-gluon exchange,
and non-perturbative, boson exchanges and confinement, aspects
of the underlying theory, QCD.
We solve the three-quark problem by means of the Faddeev method
in momentum space. We analyze the effect of the different terms in the
interaction and make contact with the use of relativistic kinematics. 
We find an explanation to the strong contribution of the
pseudoscalar forces in the semirelativistic approach for the octet baryons.
A phenomenological recipe for the regularization parameter
of the one-gluon exchange is found.
\end{abstract}

\vspace*{2cm} \noindent Keywords: \newline
nonrelativistic quark models, baryon spectra
\newline
\newline
\noindent Pacs: \newline
12.39.Jh, 12.39.Pn, 14.20.-c

\newpage

\section{Introduction}

The complexity of Quantum Chromodynamics (QCD), the quantum field theory of the
strong interaction, has prevented so far a rigorous deduction of its
predictions even for the simplest hadronic systems. In the meantime while
lattice QCD starts providing reliable results, 
QCD-inspired models are useful tools to get some insight into many of the phenomena
of the hadronic world. One of the central issues to be
addressed is a quantitative description of the low-energy phenomena,
from the baryon-baryon interaction to the baryon spectra, still
one of the major challenges in hadronic physics.

The very success of QCD-inspired models supports the picture which has
emerged from more fundamental studies, namely, that below a certain
scale QCD is a weakly coupled theory with asymptotically free quark
and gluon degrees of freedom, but above this scale a strong coupling regime
emerges in which color is confined and chiral symmetry is broken. These two
aspects, confinement and chiral symmetry breaking, are now recognized as
basic ingredients in any QCD-inspired model for the low-energy (and
therefore non-perturbative) sector. Along this line, the simplest approach
is doubtless the constituent quark model, where multigluon
degrees of freedom are eliminated in favor of confined constituent quarks
with effective masses coming from chiral symmetry breaking and quark-quark
effective interactions\cite{Man84}. 
Although little is known about the mechanism which confines the quarks
inside hadrons, unquenched lattice QCD 
suggests a linear potential screened at long-distances
due to the creation of $q\bar q$ pairs out of vacuum \cite{Bal01}.
Finally, much evidence has been accumulated about
the importance of a color-spin force (as the one 
arising from the one-gluon exchange) in
the low-energy hadron phenomenology \cite{Isg00}. 

Using these basic ingredients 
several quark models have been proposed in the
literature \cite{Val05}. In general, they were designed
either for the study of the baryon-baryon interaction or
the baryon spectra. For example,
in Refs. \cite{Shi89,Bra90,Wan92,Yuz95,Fuj04} 
the two and/or the three-nucleon problem were studied in
detail, while Refs. \cite{Isg79,Sil85,Des92,Dzi96,Glo96,Fur02} made a
thorough analysis of the baryon spectra. One of the most general conclusions
arising from these works is that
the study of a particular problem does not impose
enough restrictions as to constrain neither the ingredients
nor the parameters of the model. 
To our knowledge, in recent years the
ambitious project of a simultaneous description of the baryon-baryon
interaction and the baryon (and meson) spectra has only been undertaken by the
constituent quark model of Refs. \cite{Fer93,Gar02}, applied within the
same framework to the baryon-baryon interaction \cite{Fer93} as well as to the
baryon spectra \cite{Gar02}. This model is based on the idea that 
the constituent quark mass appears because of the spontaneous
breaking of the original chiral symmetry of the QCD Lagrangian, what
generates boson-exchange interactions between quarks.
Thus, the model takes into account perturbative and non-perturbative aspects
of QCD through the one-gluon exchange and boson exchanges and confinement,
respectively. It was originally designed 
to study the nonstrange sector and it has been recently generalized
to all flavor sectors. It has already been applied 
to the meson spectra and baryon-baryon interaction 
with encouraging results \cite{Vij04,Ter05}. 

Any phenomenological model should be tested against as many observables as possible 
to clearly understand its strengths and weakness, being this the only way one
can extract reliable predictions. 
This is why in this work we pursue the description of the nonstrange and strange
baryon spectra based on the constituent quark model of Ref. \cite{Vij04}.
For this purpose, we will start in the next section resuming its basic properties.
In Sec. III we will briefly describe the Faddeev method in momentum space
used to solve the three-body problem. Section IV will be devoted to present 
and discuss the results in comparison to other models in the literature. 
Finally, in Sec. V we will summarize our conclusions.

\section{SU(3) constituent quark model}
\label{sec2}

Let us outline the basic ingredients of the constituent quark model of
Ref. \cite{Vij04}.
Since the origin of the quark model hadrons have been considered to be
built by constituent (massive) quarks. Nowadays it is widely
recognized that the constituent quark mass appears because of the spontaneous
breaking of the original chiral symmetry of the QCD Lagrangian, what 
gives rise to boson-exchange interactions between quarks.
The quark-quark meson-exchange potentials are given by:
\begin{equation}
V_{\chi}(\vec r_{ij}) \, = \,
V_{\pi}(\vec r_{ij}) \, + \,
V_{\sigma}(\vec r_{ij}) \, + \,
V_{K}(\vec r_{ij}) \, + \,
V_{\eta}(\vec r_{ij}) \, ,
\end{equation}
each contribution given by,
\begin{eqnarray}
V_{\pi}(\vec{r}_{ij})&=&{\frac{g_{ch}^{2}}{{4\pi }}}{\frac{m_{\pi}^{2}}{{\
12m_{i}m_{j}}}}{\frac{\Lambda _{\pi}^{2}}{{\Lambda _{\pi}^{2}-m_{\pi}^{2}}}}%
m_{\pi}\left[ Y(m_{\pi}\,r_{ij})-{\frac{\Lambda _{\pi}^{3}}{m_{\pi}^{3}}}%
Y(\Lambda _{\pi}\,r_{ij})\right] (\vec{\sigma}_{i}\cdot \vec{\sigma}%
_{j})\sum_{a=1}^{3}{(\lambda _{i}^{a}\cdot \lambda _{j}^{a})}\,, \nonumber \\
V_{\sigma}(\vec{r}_{ij})&=&-{\frac{g_{ch}^{2}}{{4\pi }}}
{\frac{\Lambda_{\sigma}^{2}%
}{{\ \Lambda _{\sigma}^{2}-m_{\sigma}^{2}}}}m_{\sigma}
\left[ Y(m_{\sigma}\,r_{ij})-{\frac{%
\Lambda _{\sigma}}{m_{\sigma}}} 
Y(\Lambda _{\sigma}\,r_{ij})\right]\, , \label{zp}\\
V_{K}(\vec{r}_{ij})&=&{\frac{g_{ch}^{2}}{{4\pi }}}{\frac{m_{K}^{2}}{{\
12m_{i}m_{j}}}}{\frac{\Lambda _{K}^{2}}{{\Lambda _{K}^{2}-m_{K}^{2}}}}m_{K}%
\left[ Y(m_{K}\,r_{ij})-{\frac{\Lambda _{K}^{3}}{m_{K}^{3}}}Y(\Lambda
_{K}\,r_{ij})\right] (\vec{\sigma}_{i}\cdot \vec{\sigma}_{j})\sum_{a=4}^{7}{%
(\lambda _{i}^{a}\cdot \lambda _{j}^{a})}\, , \nonumber \\
V_{\eta }(\vec{r}_{ij})&=&{\frac{g_{ch}^{2}}{{4\pi }}}{\frac{m_{\eta }^{2}%
}{{\ 12m_{i}m_{j}}}}{\frac{\Lambda _{\eta }^{2}}{{\Lambda _{\eta
}^{2}-m_{\eta }^{2}}}}m_{\eta }\left[ Y(m_{\eta }\,r_{ij})-{\frac{\Lambda
_{\eta }^{3}}{m_{\eta }^{3}}}Y(\Lambda _{\eta }\,r_{ij})\right] (\vec{\sigma}%
_{i}\cdot \vec{\sigma}_{j})\left[ cos\theta_P(\lambda _{i}^{8}\cdot
\lambda _{j}^{8})-sin\theta _P\right] \, , \nonumber
\end{eqnarray}
the angle $\theta_P$ appears as a consequence of considering
the physical $\eta$ instead the octet one. 
$g_{ch}=m_{q}/f_{\pi }$, the $\lambda ^{\prime }s$ are the $SU(3)$
flavor Gell-Mann matrices. $m_{i}$ is the quark
mass and $m_{\pi}$, $m_{K}$ and $m_{\eta }$ are the masses of the $SU(3)$
Goldstone bosons, taken to be their experimental values. $m_{\sigma}$ is
determined through the PCAC relation 
$m_{\sigma}^{2}\sim m_{\pi}^{2}+4\,m_{u,d}^{2}$ \cite{Sca82}. 
Finally, $Y(x)$ is the standard Yukawa function defined by $Y(x)=e^{-x}/x$.

QCD perturbative effects are taken into account
through the one-gluon-exchange (OGE) potential \cite{Ruj75}.
The nonrelativistic reduction of the one-gluon-exchange diagram in QCD for point-like
quarks presents a contact term that, when not treated perturbatively, 
leads to collapse \cite{Bha80}. This is why one
maintains the structure of the OGE, but the $\delta$ function is
regularized in a suitable way. This regularization, justified by 
the finite size of the systems studied, has to be
flavor dependent \cite{Wei83}. As a consequence, the 
OGE reads,
\begin{equation}
V_{OGE}(\vec{r}_{ij}) ={\frac{1}{4}}\alpha _{s}\,\vec{\lambda ^{c}}%
_{i}\cdot \vec{\lambda^{c}}_{j}\,\left\{ {\frac{1}{r_{ij}}}-{\frac{1}{%
6m_{i}m_{j}}}\vec{\sigma}_{i}\cdot \vec{\sigma}_{j}
\,{\frac{{e^{-r_{ij}/r_{0}(\mu )}}}{r_{ij}\,
r_0^2(\mu)}}\right\} \, ,
\end{equation}
where $\lambda^{c}$ are the $SU(3)$ color matrices, 
$\alpha_s$ is the quark-gluon coupling constant, and
$r_0(\mu)=\hat r_0 \mu_{nn}/\mu_{ij}$, where $\mu_{ij}$ is the 
reduced mass of quarks $ij$ ($n$ stands for the light $u$ and $d$
quarks) and $\hat r_0$ is a parameter to be determined from the data.

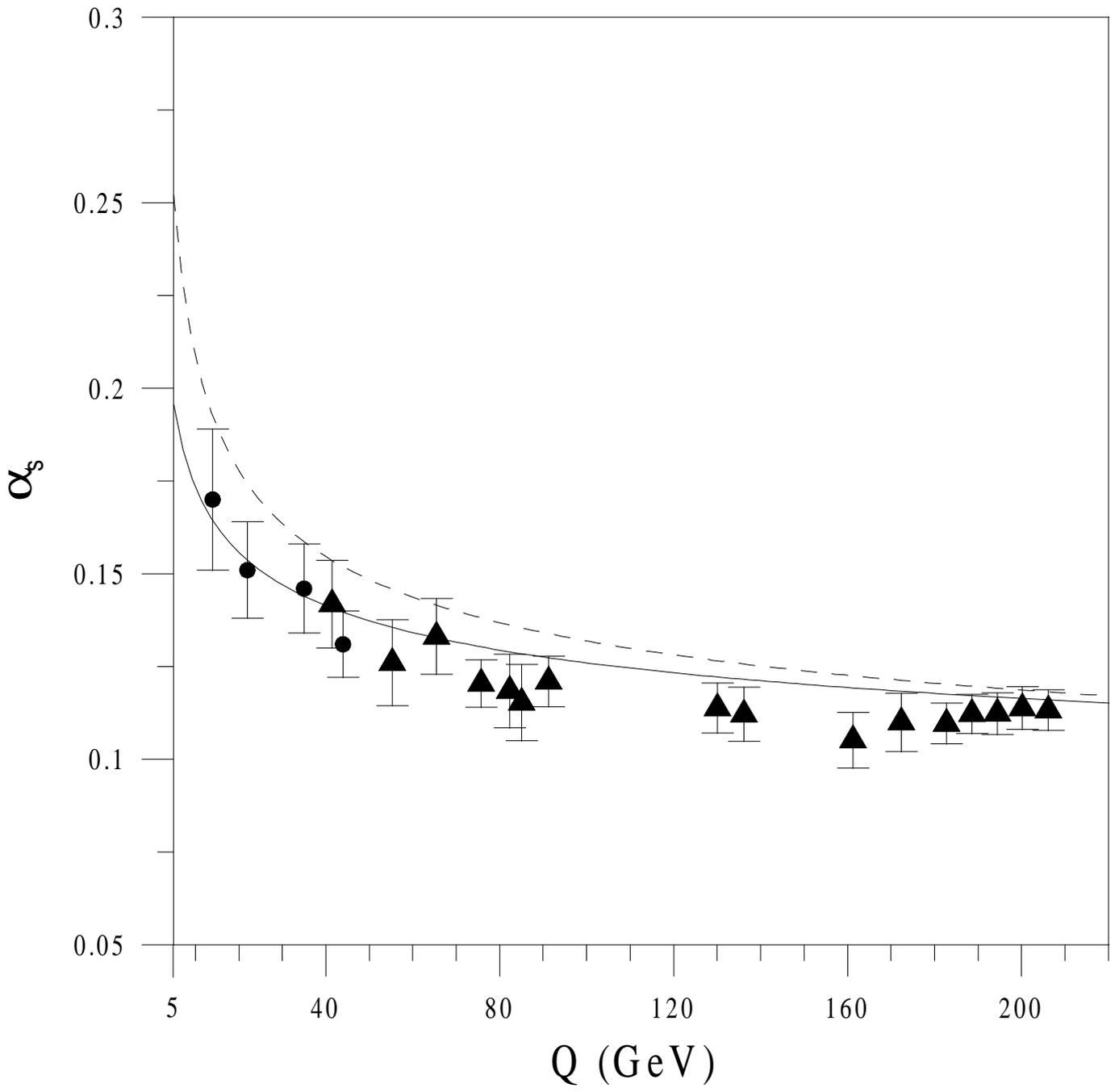
\begin{figure}
\caption{Effective scale-dependent strong coupling constant $\alpha_s$ given
in Eq. (\protect\ref{asf}) as a function of momentum. 
The solid line represents our parametrization. 
Dots and triangles are the experimental results
of Refs. \protect\cite{Klu03}.
We plot by a dashed line the parametrization obtained
in Ref. \protect\cite{Shi97} using $\Lambda=$ 0.2 GeV.}
\label{fig1}
\end{figure}

The strong coupling constant, taken to be constant
for each flavor sector, has to be scale-dependent 
when describing different flavor sectors \cite{Tit95}. 
Such an effective scale dependence has been related to the typical
momentum scale of each flavor sector assimilated
to the reduced mass of the system \cite{Hal93}.
This has been found to be relevant for the study of the
meson spectra within the present model \cite{Vij04}. 
In our case, without being a relevant parameter, we will 
respect the nice determination established there, 
\begin{equation}
\alpha_s(\mu)={\alpha_0\over{ln\left[{({\mu^2+\mu^2_0})/
\gamma_0^2}\right]}},
\label{asf}
\end{equation}
where $\mu$ is the reduced mass of the interacting $qq$ 
pair and $\alpha_0=2.118$, 
$\mu_0=36.976$ MeV and $\gamma_0=0.113$ fm$^{-1}$.
This equation gives rise to $\alpha_s\sim0.54$ for the light-quark sector,
a value consistent with the one used in the study of the nonstrange 
hadron phenomenology \cite{Fer93,Gar02}, $\alpha_s\sim0.49$ for a light-strange pair
and $\alpha_s\sim0.44$ for the strange sector, and it also has an appropriate 
high $Q^2$ behavior, $\alpha_s\sim0.127$ at the $Z_0$ mass \cite{Dav97}.
In Fig. \ref{fig1} we compare this parametrization to the experimental 
data \cite{Klu03} and to the parametrization 
obtained in Ref. \cite{Shi97} from an analytical model of QCD.

Finally, any model imitating QCD should incorporate
confinement. Lattice calculations
in the quenched approximation derived, for heavy quarks, a
confining interaction linearly dependent on the interquark
distance.  The consideration of sea quarks apart from valence 
quarks (unquenched approximation) suggests a screening effect 
on the potential when increasing the interquark distance \cite{Bal01}.
A screened potential simulating these results 
can be written as,
\begin{equation}
V_{CON}(\vec{r}_{ij})=-a_{c}\,(1-e^{-\mu_c\,r_{ij}})
(\vec{\lambda^c}_{i}\cdot \vec{ \lambda^c}_{j}) \, \, .
\end{equation}
At short distances it presents 
a linear behavior with an effective confinement strength 
$a=a_c \, \mu_c \, \vec{\lambda^c}_i \cdot \vec{\lambda^c}_j$, while
it becomes constant at large distances. Screened 
confining potentials have been analyzed in the literature
providing an explanation to the
missing state problem in the baryon spectra \cite{Vij03},
improving the description of the heavy-meson spectra \cite{Ped03},
and justifying the
deviation of the meson Regge trajectories from the linear behavior
for higher angular momentum states \cite{Bri00}.

We have not considered the noncentral contributions arising from
the different terms of the interacting potential. Experimentally, there
is no evidence for important effects of the noncentral terms on the
baryon spectra. This is clearly observed in the almost degeneracy of the
nucleon ground states with $J^\pi=1/2^-$ and $J^\pi=3/2^-$, or their
first excited states with the nucleon ground state with $J^\pi=5/2^-$. 
The same is observed around the whole
baryon spectra except for the particular problem of the
relative large separation between the $\Lambda(1405)$, $J^\pi=1/2^-$,
and the $\Lambda(1520)$, $J^\pi=3/2^-$, related to the
vicinity of the $N \overline K$ threshold \cite{Vei85}.

Theoretically, the
spin-orbit force generated by the OGE has been justified to cancel with
the Thomas precession term obtained from the confining potential \cite{Isg99}.
This is not however the case for the two-baryon system where,
by means of an explicit model for confinement, it has been demonstrated
that the strong cancellation in the baryon
spectra translates into a constructive effect 
for the two-baryon system \cite{Koi86}.
One should notice that the scalar boson-exchange potential also
presents a spin-orbit contribution with the same properties as before,
it cancels the OGE spin-orbit force in the baryon spectra
while it adds to the OGE contribution for the nucleon-nucleon $P-$waves 
and cancels for $D-$waves \cite{Val95},
as it is observed experimentally. 
Such a different behavior in the one- and two-baryon systems
is due to the absence of a direct term in the OGE spin-orbit force
(due to the color of the gluon only quark-exchange
diagrams are allowed), while the spin-orbit contribution of the
confining interaction in Ref. \cite{Koi86} and that of the
scalar boson-exchange potential in Ref. \cite{Val95} 
are dominated by a direct term, without
quark exchanges. Regarding 
the tensor terms of the meson-exchange potentials, they have been explicitly 
evaluated in the literature (in a model with stronger meson-exchange
potentials) finding contributions not bigger that 25 MeV \cite{Fur03}. 
This is due to the fact that the tensor terms give 
their most important contributions at intermediate
distances (of the order of 1-2 fm), due to the direct term in the
quark-quark potential. The regularization of the boson-exchange
potentials below the chiral symmetry breaking scale
suppresses their contributions for the very small distances
involved in the one-baryon problem.
This allows to
neglect the noncentral terms of the interacting potential
that would provide with a fine tune of the final 
results and would make very much involved and time-consuming
the solution of the three-body problem by means of the Faddeev method
in momentum space we pretend to use.

Once perturbative (one-gluon exchange) and nonperturbative (confinement
and chiral symmetry breaking) aspects of QCD have been considered, one
ends up with a quark-quark interaction of the form,
\begin{equation}
V_{q_iq_j} (\vec r_{ij}) =
V_{CON} (\vec r_{ij}) + V_{OGE} (\vec r_{ij}) + V_{\chi} (\vec r_{ij})
\label{pot}
\end{equation}

\section{Three-body formalism}

If there are no tensor or spin-orbit forces the Faddeev equations for the
bound-state problem of three quarks can be written as
\begin{eqnarray}
<p_iq_i;\ell_i\lambda_iS_iT_i|\phi_i^{LST}>  =  
{1\over E-p_i^2/2\eta_i-q_i^2/2\nu_i}
\sum_{j\ne i}
\sum_{\ell_j\lambda_jS_jT_j}
{1\over 2}\int_{-1}^1 d{\rm cos}\theta\int_0^\infty q_j^2 dq_j \nonumber \\
 \times\, t_i^{\ell_iS_iT_i}(p_i,p_i^\prime;E-q_i^2/2\nu_i) 
A_L^{\ell_i\lambda_i\ell_j\lambda_j}(p_i^\prime q_i p_j q_j) \nonumber \\
 \times  <S_iT_i|S_jT_j>_{ST}\,
<p_jq_j;\ell_j\lambda_jS_jT_j|\phi_j^{LST}>,
\label{e5c5}
\end{eqnarray}
where $S_i$ and $T_i$ are the spin and isospin of the 
pair $jk$ while $S$ and $T$ are the total spin and isospin.
$\ell_i$ ($\vec{p}_i$) is the orbital angular momentum (momentum) 
of the pair $jk$, $\lambda_i$ ($\vec{q}_i$)
is the orbital angular momentum (momentum) 
of particle $i$ with respect to the pair
$jk$, and $L$ is the total orbital angular momentum.
${\rm cos}\theta=\vec q_i\cdot\vec q_j/(q_iq_j)$ while
\begin{eqnarray}
\eta_i & = & {m_j m_k \over m_j + m_k} \, , \nonumber \\
\nu_i & = & {m_i(m_j+m_k) \over m_i+m_j+m_k},
\end{eqnarray}
are the usual reduced masses.
For a given set of values of $LST$ the integral equations (\ref{e5c5})
couple the amplitudes of the different configurations
$\{\ell_i\lambda_i S_i T_i\}$.
The spin-isospin recoupling coefficients $<S_iT_i|S_jT_j>_{ST}$ are given by
\begin{eqnarray}
<S_iT_i|S_jT_j>_{ST}  = 
(-)^{S_j+\sigma_j-S}\sqrt{(2S_i+1)(2S_j+1)} \, W(\sigma_j\sigma_kS\sigma_i;S_iS_j)
 \nonumber \\
 \times 
(-)^{T_j+\tau_j-T}\sqrt{(2T_i+1)(2T_j+1)} \, W(\tau_j\tau_kT\tau_i;T_iT_j),
\label{e8c5}
\end{eqnarray}
with $\sigma_i$ and $\tau_i$ the spin and isospin of particle $i$, and $W$
is the Racah coefficient.
The orbital angular momentum recoupling coefficients 
$A_L^{\ell_i\lambda_i\ell_j\lambda_j}(p_i^\prime q_i p_j q_j)$ are given by 
\begin{eqnarray}
A_L^{\ell_i\lambda_i\ell_j\lambda_j}(p_i^\prime q_i p_j q_j)  = 
{1\over 2L+1}\sum_{M m_i m_j}C^{\ell_i \lambda_i L}_{m_i,M-m_i,M}
C^{\ell_j \lambda_j L}_{m_j,M-m_j,M}\Gamma_{\ell_i m_i}\Gamma_{\lambda_i
M-m_i}\Gamma_{\ell_j m_j} \nonumber \\  \times
\Gamma_{\lambda_j M-m_j}
{\rm cos}[-M(\vec q_j,\vec q_i)-m_i(\vec q_i,{\vec p_i}^{\,\prime})
+m_j(\vec q_j,\vec p_j)],
\label{e9c5}
\end{eqnarray}
with $\Gamma_{\ell m}=0$ if $\ell -m$ is odd and
\begin{equation}
\Gamma_{\ell m}={(-)^{(\ell+m)/2}
\sqrt{(2\ell+1)(\ell+m)!(\ell-m)!}
\over 2^\ell((\ell+m)/2)!((\ell-m)/2)!}
\label{e10c5}
\end{equation}
if $\ell-m$ is even. The angles 
$(\vec q_j,\vec q_i)$, $(\vec q_i,{\vec p_i}^{\,\prime})$, and
$(\vec q_j,\vec p_j)$ can be obtained in terms of the magnitudes of the 
momenta by using the relations
\begin{eqnarray}
\vec p_i^{\,\prime} &=& - \vec q_j - {\eta_i\over m_k}\vec q_i \, , \nonumber \\
\vec p_j &=& \vec q_i + {\eta_j\over m_k} \vec q_j,
\label{e11c5}
\end{eqnarray}
where $ij$ is a cyclic pair.
The magnitude of the momenta $p_i^\prime$ and $p_j$, on the other hand,
are obtained in terms of $q_i$, $q_j$, and
${\rm cos}\theta$ using Eqs. (\ref{e11c5}) as 
\begin{eqnarray}
p_i^\prime &=& \sqrt{q_j^2+
\left({\eta_i\over m_k}\right)^2q_i^2
+{2\eta_i\over m_k}q_i q_j {\rm cos}\theta}, \nonumber \\
p_j &=& \sqrt{q_i^2+
\left({\eta_j\over m_k}\right)^2q_j^2
+{2\eta_j\over m_k}q_i q_j {\rm cos}\theta}.
\end{eqnarray}

Finally, the two-body amplitudes 
$t_i^{\ell_iS_iT_i}(p_i,p_i^\prime;E-q_i^2/2\nu_i)$
are given by the solution of the Lippmann-Schwinger equation
\begin{eqnarray}
t_i^{\ell_iS_iT_i}(p_i,p_i^\prime;E-q_i^2/2\nu_i)  = 
V_i^{\ell_iS_iT_i}(p_i,p_i^\prime) + \int_0^\infty {p_i^{\prime\prime}}^2 
dp_i^{\prime\prime}\, V_i^{\ell_iS_iT_i}(p_i,p_i^{\prime\prime})
 \nonumber \\  \times
{1\over E-{p_i^{\prime\prime}}^2/2\eta_i-q_i^2/2\nu_i}\,
 t_i^{\ell_iS_iT_i}(p_i^{\prime\prime},p_i^\prime;E-q_i^2/2\nu_i),
\label{e15c5}
\end{eqnarray}
with
\begin{equation}
V_i^{\ell_iS_iT_i}(p_i,p_i^\prime) 
= {2\over \pi}\int_0^\infty r_i^2 dr_i\,j_{\ell_i}(p_ir_i)
V_i^{S_iT_i}(r_i)j_{\ell_i}(p_i^\prime r_i).
\label{e16c5}
\end{equation}
and $j_{\ell}$ the spherical Bessel function.

In the case where the three quarks are identical ($N$ and $\Omega$) the 
three amplitudes $\phi_1^{LST}$, $\phi_2^{LST}$, and $\phi_3^{LST}$ 
in Eq. (\ref{e5c5}) are identical so that it reduces to
\begin{eqnarray}
<p_iq_i;\ell_i\lambda_iS_iT_i|\phi^{LST}>  =  
{1\over E-p_i^2/2\eta_i-q_i^2/2\nu_i}
\sum_{\ell_j\lambda_jS_jT_j}
\int_{-1}^1 d{\rm cos}\theta\int_0^\infty q_j^2 dq_j \nonumber \\
 \times\, t_i^{\ell_iS_iT_i}(p_i,p_i^\prime;E-q_i^2/2\nu_i) 
A_L^{\ell_i\lambda_i\ell_j\lambda_j}(p_i^\prime q_i p_j q_j) \nonumber \\
 \times  <S_iT_i|S_jT_j>_{ST}\,
<p_jq_j;\ell_j\lambda_jS_jT_j|\phi^{LST}> \, \, ,
\label{e17c5}
\end{eqnarray}
with $(-)^{\ell_i+S_i+T_i}=1$ as required by the
Pauli principle since the wave function is color antisymmetric.

In the case where two quarks are identical and one is different ($\Lambda$,
$\Sigma$, and $\Xi$) only two amplitudes are independent. Assuming that 
particles 2 and 3 are identical and 1 is different, only the amplitudes
$\phi_1^{LST}$ and $\phi_2^{LST}$ are independent and satisfy the coupled
integral equations \cite{Afn74,Gar90}
\begin{eqnarray}
<p_2q_2;\ell_2\lambda_2S_2T_2|\phi_2^{LST}>  = G
{1\over E-p_2^2/2\eta_2-q_2^2/2\nu_2}
\sum_{\ell_3\lambda_3S_3T_3}
{1\over 2}\int_{-1}^1 d{\rm cos}\theta\int_0^\infty q_3^2 dq_3 \nonumber \\
 \times\, t_2^{\ell_2S_2T_2}(p_2,p_2^\prime;E-q_2^2/2\nu_2) 
A_L^{\ell_2\lambda_2\ell_3\lambda_3}(p_2^\prime q_2 p_3 q_3) \nonumber \\
 \times  <S_2T_2|S_3T_3>_{ST}\,
<p_3q_3;\ell_3\lambda_3S_3T_3|\phi_2^{LST}> \nonumber \\
+{1\over E-p_2^2/2\eta_2-q_2^2/2\nu_2}
\sum_{\ell_1\lambda_1S_1T_1}
{1\over 2}\int_{-1}^1 d{\rm cos}\theta\int_0^\infty q_1^2 dq_1 \nonumber \\
 \times\, t_2^{\ell_2S_2T_2}(p_2,p_2^\prime;E-q_2^2/2\nu_2) 
A_L^{\ell_2\lambda_2\ell_1\lambda_1}(p_2^\prime q_2 p_1 q_1) \nonumber \\
 \times  <S_2T_2|S_1T_1>_{ST}\,
<p_1q_1;\ell_1\lambda_1S_1T_1|\phi_1^{LST}>,
\label{e18c5}
\end{eqnarray}
\begin{eqnarray}
<p_1q_1;\ell_1\lambda_1S_1T_1|\phi_1^{LST}>  =  
{1\over E-p_1^2/2\eta_1-q_1^2/2\nu_1}
\sum_{\ell_2\lambda_2S_2T_2}
\int_{-1}^1 d{\rm cos}\theta\int_0^\infty q_2^2 dq_2 \nonumber \\
 \times\, t_1^{\ell_1S_1T_1}(p_1,p_1^\prime;E-q_1^2/2\nu_1) 
A_L^{\ell_1\lambda_1\ell_2\lambda_2}(p_1^\prime q_1 p_2 q_2) \nonumber \\
 \times  <S_1T_1|S_2T_2>_{ST}\,
<p_2q_2;\ell_2\lambda_2S_2T_2|\phi_2^{LST}>,
\label{e19c5}
\end{eqnarray}
where the identical-particles phase $G$ is
\begin{equation}
G=(-1)^{1+\ell_2+\sigma_1+\sigma_3-S_2+\tau_1+\tau_3-T_2}.
\label{e20c5}
\end{equation}
Substituting Eq. (\ref{e19c5}) into Eq. (\ref{e18c5}) one obtains a
single integral equation for the amplitude $\phi_2^{LST}$.
Again, in the case of identical pairs one has
$(-)^{\ell_1+S_1+T_1}=1$.

The nonrelativistic Faddeev method has a problem if the two-body
interactions allow transitions of the form $a + b \to c + d$, i.e., if
the particles in the final state are different from the ones in the
initial state. In that case the center of mass energy is different in the initial
and final states. This problem, however, does not arise in our model
since our two-body interactions given by Eq. (\ref{pot}) only allow
transitions of the form $n + n \to n + n$, $n + s \to n + s$, and
$s + s \to s + s$, $n$ standing for a light $u$ or $d$ quark. 
The center of mass ambiguity in the case of transitions of
the form $a + b \to c + d$ does not arise in the relativistic version
of the Faddeev method described in Ref. \cite{Gar03b}.

\section{Results and discussion}

The results we are going to present have been obtained by solving
exactly the Schr\"odinger equation by the Faddeev method in momentum
space we have just described. For baryons made up of three 
identical quarks we have also calculated the spectra by means of 
the hyperspherical harmonic (HA) expansion method \cite{Bar00}.
The HA treatment allows a more intuitive understanding of the wave functions in
terms of the hyperradius of the whole system. These wave functions
will be used to calculate the root mean square radius. 
As a counterpart one has to go to a very high order in the expansion 
to get convergence. To assure this we shall expand up to $K=24$ ($K$ being 
the great orbital determining the order of the expansion).
Differences in the results for the $3q$ bound state
energies obtained by means of the two methods turn out to be at most of 5 MeV.

As mentioned above we will not perform a systematic study in order
to determine the best set of parameters to fit the 
baryon spectra. Instead, we will start from the 
parameters used in Ref. \cite{Vij04} for the description 
of the meson spectra that are resumed in Table \ref{t1}. 
There are two parameters that may differ from the meson case, they are: 
$\hat r_0$, connected to the typical size of the system where 
the contact interaction is regularized and $a_c$, the strength of
confinement. We fix $a_c$ to drive the Roper of the nucleon to its
correct position. One could also have chosen to fix the negative 
parity states knowing the sensitivity of the Roper resonance to 
the kinematics used \cite{Car83,Gar03}, however we prefer 
to maintain the same prescription as in the study
of the nonstrange baryon spectra \cite{Gar02}, to guarantee
that a similar description is obtained for the light baryons.
We fix $\hat r_0$ to have the correct $\Delta -N$ mass difference. 
Once we determine $\hat r_0$ for the light 
baryons, its value is determined for all other 
flavor sectors through the relation given in Sec. \ref{sec2}, 
obtaining a correct description of all hyperfine splittings.
Finally, we made a fine tune of the strange quark mass to improve
the description of the ground states with 
strangeness different from zero.

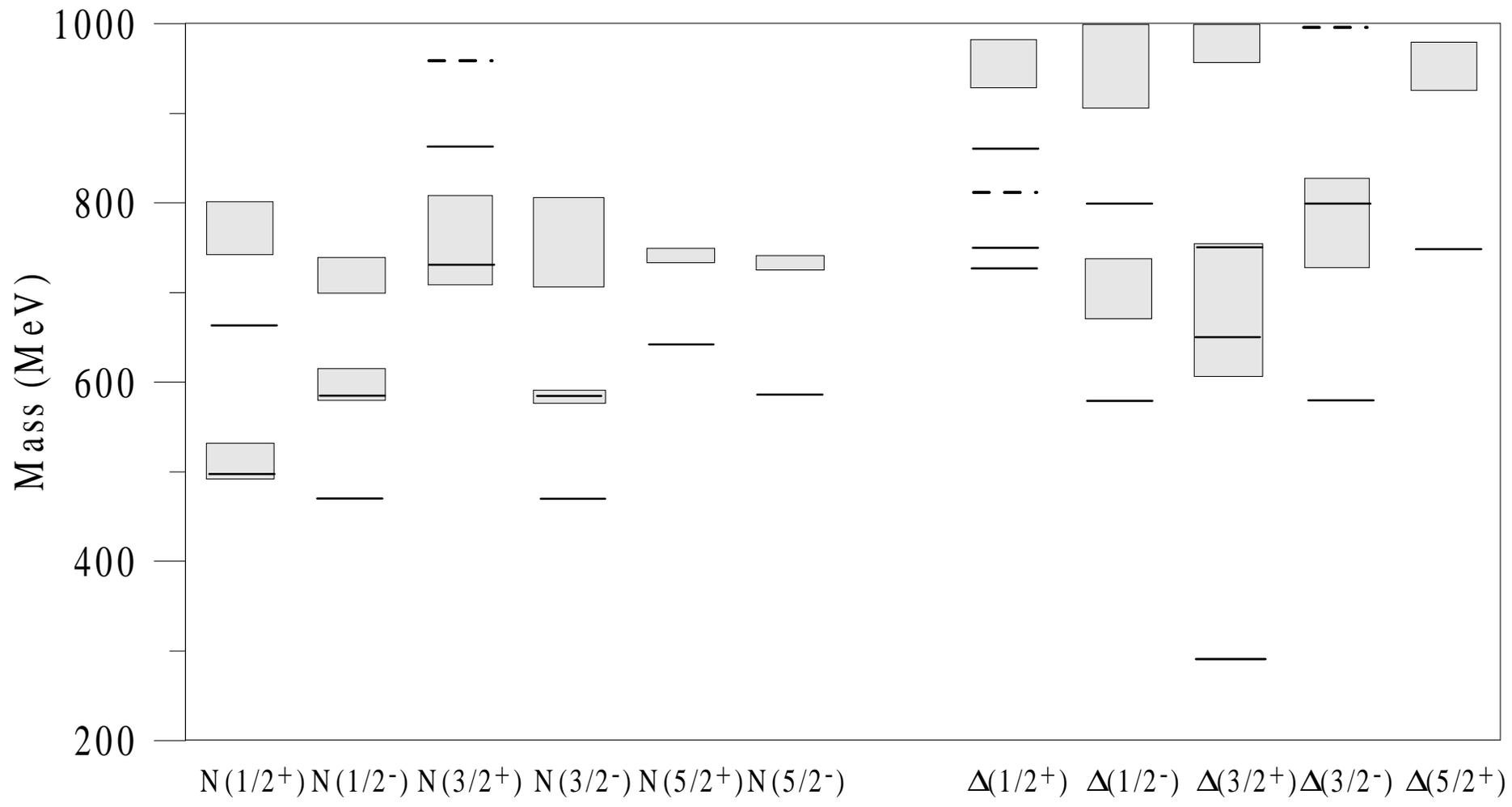
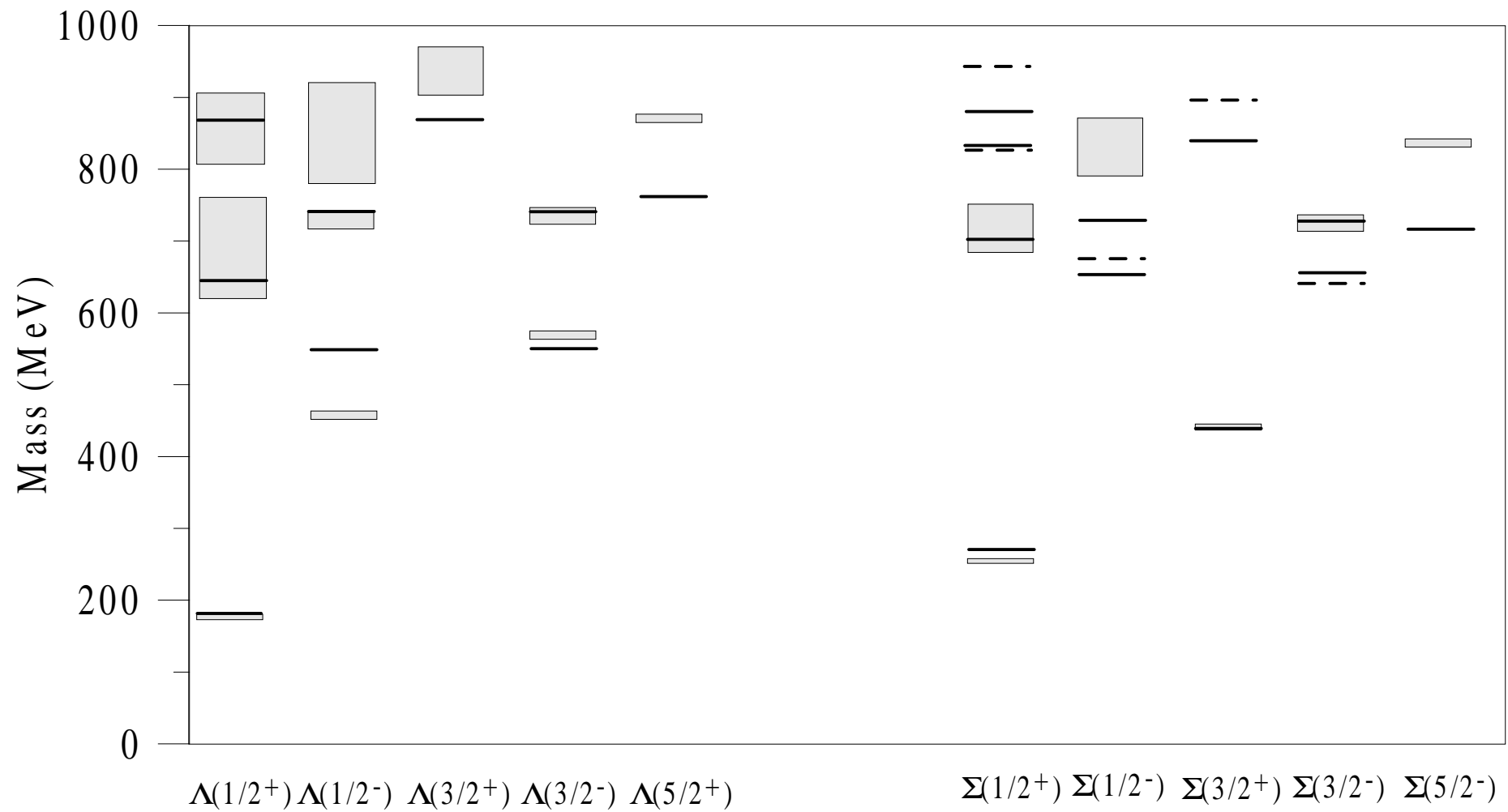
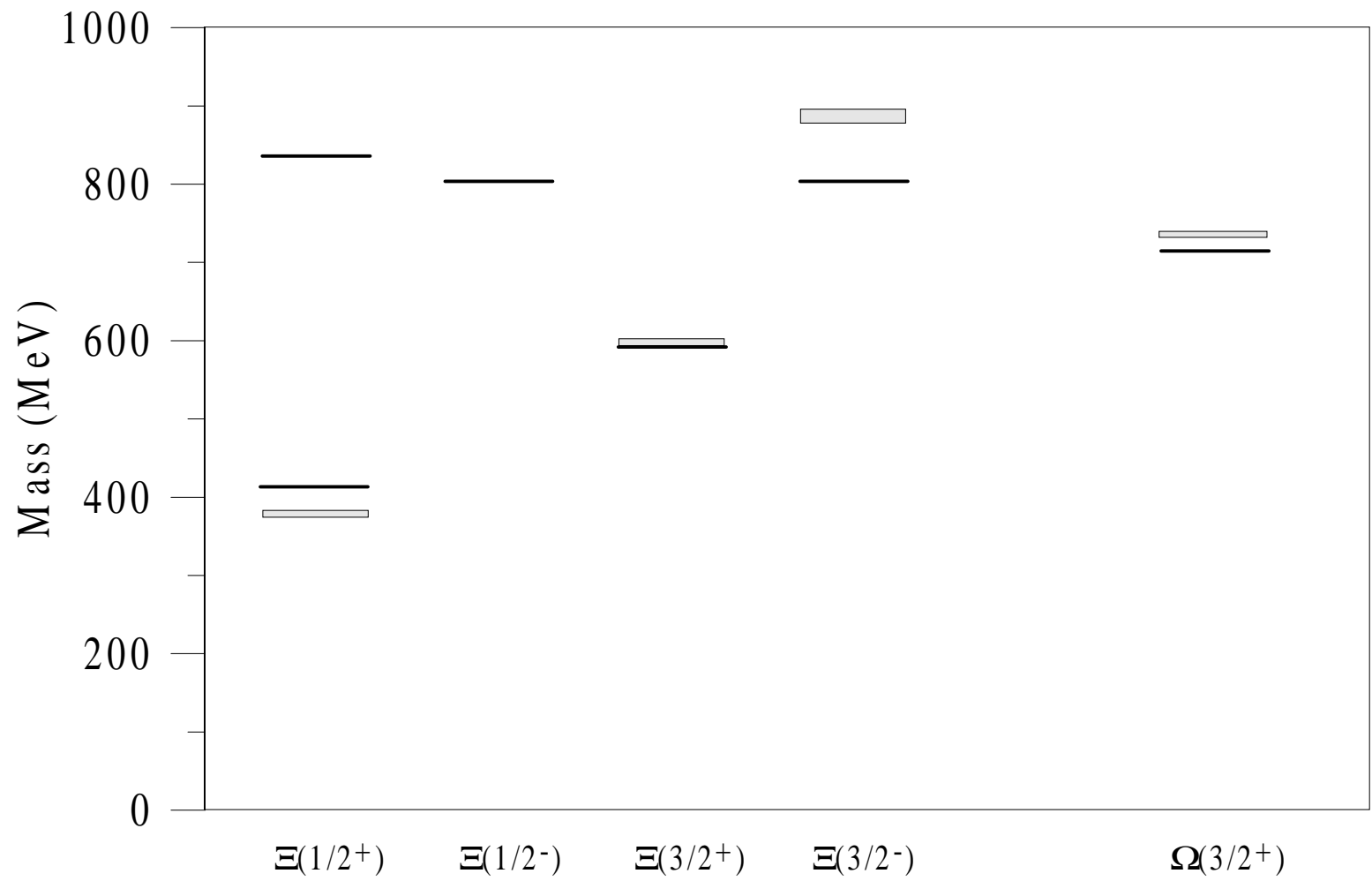
\begin{figure}[tbp]
\caption{ Relative energy (a) $N$ and $\Delta$, (b) $\Lambda$
and  $\Sigma$, (c) $\Xi$ and
$\Omega$ spectra up to 1.0 GeV excitation energy. The solid lines
correspond to the results of our  model. The shaded regions, whose size
stands for the experimental uncertainty, represent the experimental
data. The dashed lines stand for experimental states whose 
mass is given but without indicating the error bars.}
\label{fig2}
\end{figure}

Our results are shown in Fig. \ref{fig2} for the different
octet and decuplet baryons. 
As can be seen our election of fixing $a_c$ to reproduce
the Roper resonance gives, in general, masses somewhat smaller
than experiment. As explained above, we could equally have determined $a_c$ to describe
the negative parity states producing a much better fit of the baryon spectra
except for the Roper resonance, that it is know to decrease in energy
when a semirelativistic prescription is used \cite{Gar03}.
Let us focus our attention on several particular
aspects that deserve a detailed discussion. A widely discussed issue
on the baryon spectra has been the so-called level ordering
problem. It can be easily illustrated for the nucleon spectrum
in the pure harmonic limit. The $N^*(1440)$ $J^P=1/2^+$ belongs to the $[56,0^+]$
$SU(6)_{FS} \times O(3)$ irreducible representation and it appears in the $N=2$ band,
while the $N^*(1535)$ $J^P=1/2^-$ belongs to the
$[70,1^-]$ appearing in the $N=1$ band. As a consequence,
the $N^*(1440)$ has $2 \hbar \omega$ energy excitation while the
$N^*(1535)$ has only $1 \hbar \omega$ energy excitation,
opposite to the order observed experimentally. Theoretically,
this situation has been cured by means of appropriate phenomenological
interactions as it is the case of anharmonic terms \cite{Isg00},
scalar three-body forces \cite{Des92}, or pseudoscalar interactions
\cite{Glo96,Gar02}.

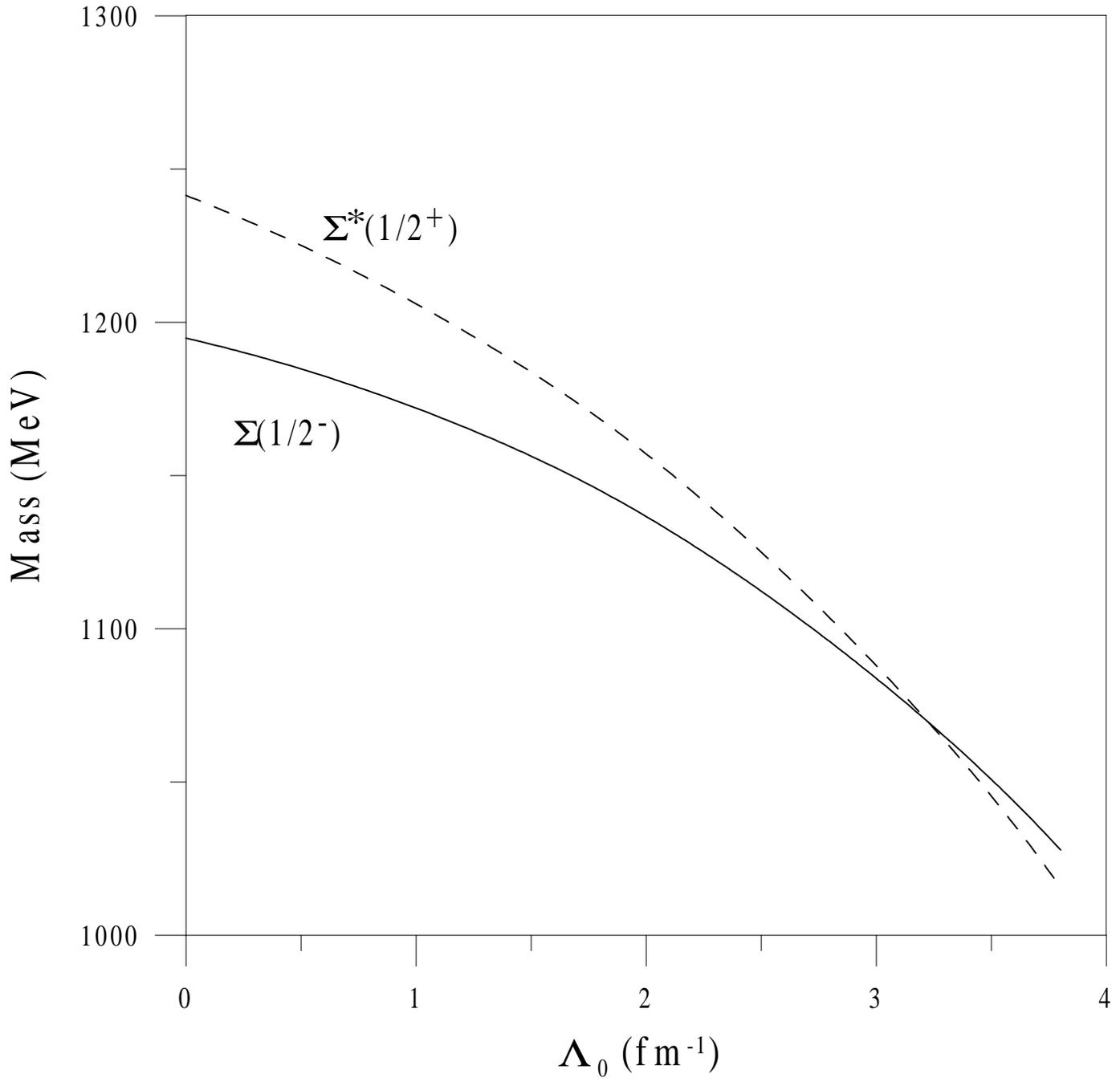
\begin{figure}[tbp]
\caption{$\Sigma^*(1/2^+)$ and $\Sigma(1/2^-)$ masses as a function of the
cutoff masses of the one-pion and one-kaon exchanges,
$\Lambda'_{\pi,K}=\Lambda_{\pi,K}+\Lambda_0$.} 
\label{fig3}
\end{figure}

The mechanism producing the reverse of the ordering between the 
positive and negative parity excited states is the following.
In the case of the 
scalar three-body force of Ref. \cite{Des92}, in the limit of zero range it 
would act only for states whose wave function do not cancel at the origin.
It therefore influences the $L=0$ ground states and their radial
excitations, while producing essentially no effect for states with
mixed symmetry (negative parity states). As a consequence,
if this force is chosen attractive, it explains
why the Roper resonances are lower than the negative
parity excited states.
In the case of the chiral pseudoscalar interaction, its
$(\vec{\sigma} \cdot \vec{\sigma}) (\vec{\lambda} \cdot \vec{\lambda})$
structure gives attraction for symmetric spin-flavor pairs and
repulsion for antisymmetric ones.
This lowers the position of the first radial excitation, 
with a completely symmetric spin-flavor wave function,
with regard to the first negative parity state, with a 
spin-flavor mixed symmetry wave function. 
This effect appears in our model mainly through
the one-pion and one-kaon exchange contributions. It has been 
illustrated in Fig. \ref{fig3}, where we plot
the mass of the first radial and orbital excitations 
of the $\Sigma(1/2^+)$ as a function of the cutoff mass
of the one-pion and one-kaon exchange potentials. 
The contribution of the pseudoscalar interactions is increased by
letting the cutoff parameters $\Lambda_{\pi ,K}$ to grow
in the same manner $\Lambda_{\pi ,K}'=\Lambda_{\pi ,K}+\Lambda_0$.
As can be seen, the reverse of
the ordering between the positive and negative parity 
excited states is obtained for $\Lambda_0$ sufficiently large
(around 3.2 fm$^{-1}$, $\Lambda_\pi=7.4$ fm$^{-1}$ and
$\Lambda_K=8.4$ fm$^{-1}$). 
A model with such a strong cutoffs
would not be realistic because the decuplet-octet [$\Sigma(3/2^+)-\Sigma(1/2^+)$]
mass difference would be much larger than the experimental value.
This difficulty is known to have a well defined and simple
solution, due to the decreasing of the
excitation energy of the nucleon Roper resonance induced by the relativistic
kinematics \cite{Car83,Gar03}, which would reduce the value of the
cutoff needed. 

Although the level ordering problem has been solved by potential models 
based only on pseudoscalar forces combined with relativistic kinematics,
they give rise to very small sizes for baryons. 
We compare in Table \ref{t3} the root mean
square radii obtained with the constituent quark model 
used in this work to those of Ref. \cite{Des92},
making use of a scalar three-body force, Ref. \cite{Fur02}, based only
on pseudoscalar forces and relativistic kinematics, and Ref. \cite{Sil85} 
based on the Bhaduri potential. Ref. \cite{Des92} gives a very small size 
for the nucleon while Ref. \cite{Fur02} finds small sizes for all baryons.
The model based on the Bhaduri potential \cite{Sil85} produces 
sizes closer to our model. These results can be
understood in the following way.
As explained above, the scalar three-body force of Ref. \cite{Des92}
gives a strong attraction for the nucleon and its radial excitations,
being the responsible for their small radius, while it produces practically
no effect on the other baryons, being their radius
much bigger and comparable to those of Ref. \cite{Sil85}.
In Ref. \cite{Fur02}, the contribution of the pseudoscalar 
boson exchanges to the baryon masses (see Table II of Ref. \cite{Fur02})
is very large, specially for the octet baryons, being the responsible for their
small sizes. Although for the decuplet baryons 
this contribution is reduced, the sizes obtained are still very small. 
As we will explain below this is a direct consequence of 
smearing out the pseudoscalar meson exchange delta function
with a large cutoff. This is reflected, for example, in the
mass difference induced by the one-pion and one-kaon exchanges
between decuplet and octet
baryons, $\Delta -N$, $\Sigma (3/2^+)-\Sigma (1/2^+)$, $\Xi (3/2^+)-\Xi (1/2^+)$,
of the order of 900 MeV. 

In the constituent quark model used in
this work the hyperfine splitting is shared between
pseudoscalar forces and perturbative QCD contributions, 
provided by the one-gluon exchange. 
In Table \ref{t2} we give the contribution of different pieces of the
interacting hamiltonian to the energy of several octet and decuplet baryons.
One observes that the hyperfine splittings are
basically controlled by the OGE ($V_2$) 
and OPE ($V_3$) [OKE ($V_5$)] potentials in the non-strange [strange] sector.
When the OGE and OPE are considered altogether ($V_4$) the splitting 
is bigger that the sum of both contributions separately, and 
they generate almost the experimental hyperfine splitting, the $\eta$ and
$\sigma$ given a final small tune.
The expectation value of the OPE flavor operator for two light quarks,

\begin{equation}
\left \langle [f_{ij}]_F \,\, T_{ij} \right|  \sum_{a=1}^{3} 
\lambda_i^a \lambda_j^a \left|
[f_{ij}]_F \,\, T_{ij} \right \rangle = \left\{\matrix{1 & & {\rm if} 
\,\,\,\,\,\, [2]_F, T_{ij}=1 \cr
-3 & & {\rm if} \,\,\,\,\,\, [11]_F, T_{ij}=0} \right.
\end{equation}
is replaced by the similar effect of the OKE when a light and a strange quarks are involved
\begin{equation}
\left \langle [f_{ij}]_F \,\, T_{ij} \right|  \sum_{a=4}^{7} 
\lambda_i^a \lambda_j^a \left|
[f_{ij}]_F \,\, T_{ij} \right \rangle = \left\{\matrix{2 & & {\rm if} 
\,\,\,\,\,\, [2]_F, T_{ij}=1/2 \cr
-2 & & {\rm if} \,\,\,\,\,\, [11]_F, T_{ij}=1/2} \right.
\end{equation}
being $[f_{ij}]_F$ the flavor permutational symmetry in the quark pair
$(i,j)$ and $T_{ij}$ the total isospin of the pair state. 
They enhance in a similar way the hyperfine splitting
produced by the OGE: the OPE for light quark pairs and the OKE for light-strange ones.
The important effect of the OGE is observed when Table \ref{t2} is compared 
to Table II of Ref. \cite{Fur02},
the contribution of the pseudoscalar forces is much smaller
in our case,
generating decuplet-octet mass differences of the order of
100$-$200 MeV, the remaining mass difference given by the OGE.
As a consequence the radii predicted are also bigger.

This regularization effect of the OGE over the
pseudoscalar forces for the baryon spectra has been also observed in 
two-baryon calculations \cite{Nak00} (that we consider should be
proximately linked to the one-body problem). 
A too strong nucleon-nucleon pseudoscalar force was found for models based only 
in Goldstone boson exchanges and, at the same
time, they do not present the required attraction to reproduce the experimental
data \cite{Nak00}. The consideration of the scalar octet of Goldstone bosons \cite{Bar01}
(as proposed long ago in the first work of 
Ref. \cite{Fer93}) may remedy the situation for the two-body
sector, but it is incompatible with the description of the baryon spectra,
because it makes the system to collapse \cite{Fur02}. The reason for that
can be easily understood looking at the results of
Ref. \cite{Gar03b}, where it has been demonstrated that a different 
regularization scale is obtained for the same interaction when 
nonrelativistic or relativistic kinematics are used.
A larger value of the regularization parameter 
of the OGE delta function
was obtained for the case of the semirelativistic calculation (see
Fig. 1 of Ref. \cite{Gar03b}). Therefore,
the regularization process of any delta
function (as the ones present in the Goldstone boson exchanges) 
should be done with great care. 
The semirelativistic kinematics cannot be implemented
without worrying about the corrections
to the meson-exchange potential in a consistent way. 
Replacing the
nonrelativistic by the semirelativistic kinematics, the value
of the delta-function regularization parameter giving rise to unstable results 
is increased, the other way around, for the same regularization parameter the 
interaction is made much more stronger. 
In the presence of so a strong pseudoscalar force, as shown in the
results of Ref. \cite{Fur02}, the additional attraction
provided by the scalar potential gives rise to collapse. 
This is not again the case of our model where the scalar 
interaction is crucial to understand simultaneously the one and the 
two-baryon problems
and its strength is compatible
with the description of both sectors \cite{Gar03}, the one-gluon
exchange being basic for these results. The same conclusion was obtained
for the light baryons when the semirelativistic prescription was 
used \cite{Gar03}. 

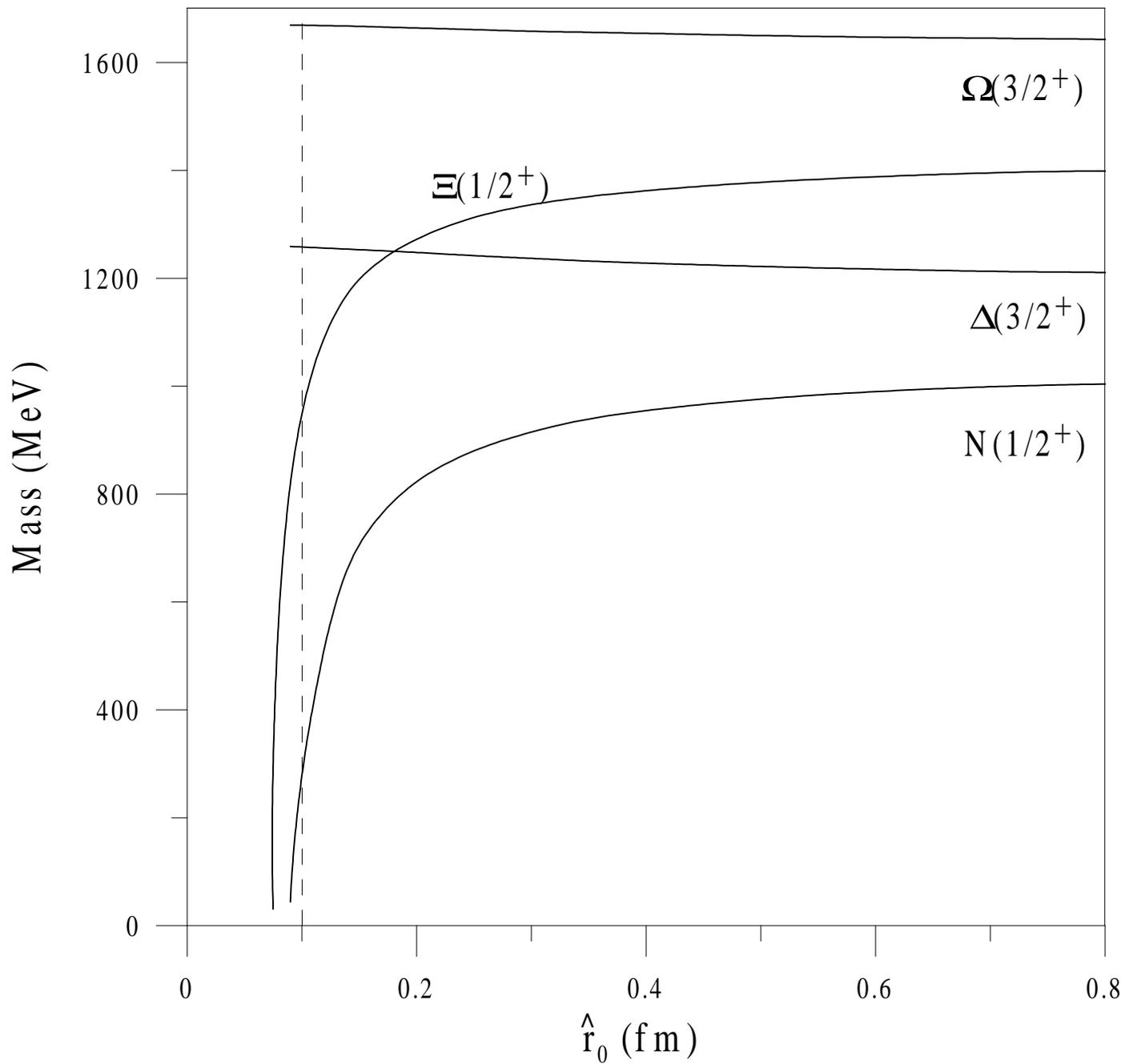
\begin{figure}[tbp]
\caption{$N(1/2^+)$, $\Xi(1/2^+)$, $\Delta(3/2^+)$ and $\Omega(3/2^+)$
ground state masses as a function of the regularization parameter $\hat r_0$.}
\label{fig4}
\end{figure}

Let us finally face the problem of the 
regularization parameter of the OGE, $r_0$. As explained
in Sec. \ref{sec2} this parameter is taken to be flavor dependent, scaling with the
reduced mass of the interacting quarks. The larger the system (the lighter the masses
of the quarks involved) the larger the value of $r_0$ that can be used without risk
of collapse. In Fig. \ref{fig4} we plot the mass of two $1/2^+$ ground states, $N$ and $\Xi$,
and two $3/2^+$ ground states, $\Delta$ and $\Omega$, as a function of $\hat r_0$. 
In the last two cases the completely
symmetric spin-flavor wave function makes the OGE to be repulsive and therefore no 
important effect is observed independently of the flavor quark substructure. However, for 
the $1/2^+$ ground states
the OGE gives attraction and the regularization should be done with care. We observe how
the masses of the $N(1/2^+)$ and the $\Xi(1/2^+)$ start to decrease very rapidly for
almost the same value of $\hat r_0$ (for $\hat r_0=0.1$ fm, marked as a vertical dashed line
in the figure, both states have diminished around 500$-$600 MeV with respect to their
asymptotic value). One should note that the value of $r_0$ for pairs 
containing strange quarks is much smaller, for example $\hat r_0=0.35$ fm implies
$r_0^{nn}=0.35$ fm while $r_0^{ns}=0.28$ fm and $r_0^{ss}=0.22$ fm. This 
flavor dependence combined with the effect of the pseudoscalar forces
provides with a correct description of the hyperfine splittings, 
giving confidence to the election of the flavor dependence of the OGE regularization
parameter. 

\section{Summary}

We have used a constituent quark model incorporating the basic properties
of QCD to study the strange and nonstrange baryon spectra. 
The model takes into account the most
important QCD nonperturbative effects: chiral symmetry breaking
and confinement as dictated by unquenched lattice QCD. It also
considers QCD perturbative effects trough a flavor dependent one-gluon exchange
potential. The parameters of the model are mostly fixed
from other observables as the meson spectra or the baryon-baryon 
interaction.

The SU(3) three-body problem has been for the first time
exactly solved by means of the
Faddeev method in momentum space, obtaining
results of similar quality to others present 
in the literature based on models specifically designed for the
study of the baryon spectra. The model
provides with baryon root mean square radii much bigger than models based only
in pseudoscalar boson exchanges. This is a consequence of the reduced 
contribution of the pseudoscalar forces due to the presence of the one-gluon
exchange. These pseudoscalar forces are important for the correct
position of the positive and negative parity excited states in all flavor sectors,
but they should not be artificially strengthened making the systems
highly unstable. The Roper resonances are know to be sensitive to
relativistic kinematics, and therefore a reduced contribution
of the pseudoscalar forces should be enough to solve the so-called
level ordering problem. The presence of the scalar Goldstone boson
exchanges, crucial to make contact with the two-body problem, would
not be compatible with a strong pseudoscalar contribution.

We have analyzed the dependence of the spectra on the regularization 
parameter of the OGE, obtaining a pretty good agreement with a scale
dependence based on the reduced mass of the interacting quarks.
This OGE potential gives an important contribution to the 
decuplet-octet mass difference being basic to regularize the 
pseudoscalar forces needed.

Finally, although we do not believe that explanations
based on constituent quark models may rule out or contradict
other alternative ones, one should acknowledge the capability
of constituent quark models 
for a coherent understanding of the low-energy phenomena of the
baryon spectroscopy and the baryon-baryon interaction in a simple 
framework based on the contribution of pseudoscalar, scalar
and one-gluon-exchange forces between quarks.

\section{acknowledgments}

This work has been partially funded by Ministerio 
de Ciencia y Tecnolog{\'{\i}}a under Contract No. FPA2004-05616,
by Junta de Castilla y Le\'{o}n under Contract No. SA-104/04,
and by COFAA-IPN (M\'exico).

\begin{table}[tbp]
\caption{Quark-model parameters.}
\label{t1}
\begin{center}
\begin{tabular}{cc|ccc}
&&$m_u=m_d$ (MeV) & 313 & \\ 
&Quark masses&$m_s$ (MeV)     & 500 & \\ 
\hline
&&$m_{\pi}$ (fm$^{-1}$) & 0.70&\\
&&$m_{\sigma}$ (fm$^{-1}$)& 3.42&\\ 
&&$m_{\eta}$ (fm$^{-1}$)  & 2.77&\\ 
&Goldstone bosons&$m_K$ (fm$^{-1}$)       & 2.51&\\ 
&&$\Lambda_{\pi}=\Lambda_{\sigma}$ (fm$^{-1}$) & 4.20 &\\
&&$\Lambda_{\eta}=\Lambda_K$ (fm$^{-1}$) & 5.20&\\ 
&& $g_{ch}^2/(4\pi)$      & 0.54&\\ 
&&$\theta_P(^o)$          & $-$15&\\ 
\hline
&&$a_c$ (MeV)             &230&\\
&Confinement&$\mu_c$ (fm$^{-1}$)&0.70&\\ 
\hline
&OGE&$\hat r_0$ (fm)     &0.35&\\
\end{tabular}
\end{center}
\end{table}

\begin{table}[tbp]
\caption{Root mean square radii, $\left \langle{r^2} \right \rangle^{\scriptsize 1/2}$ in fm,
of states of identical particles obtained with our model (CCQM), compared
to Ref. \protect\cite{Des92}, considering a scalar three-body force,
Ref. \protect\cite{Fur02}, based only on pseudoscalar
boson exchanges and relativistic kinematics,
and Ref. \protect\cite{Sil85} based on a OGE potential.}
\label{t3}
\begin{center}
\begin{tabular}{cccccc}
State & CCQM & Ref. \protect\cite{Des92} & Ref. \protect\cite{Fur02} & Ref. \protect\cite{Sil85} \\
\hline
$N(1/2^+)$          & 0.482 & 0.38 & 0.304 & 0.467 \\
$N^*(1/2^+)$        & 0.961 & 0.79 & 0.463 & $-$   \\
$N(1/2^-)$          & 0.829 & 0.78 & $-$   & $-$   \\
$\Delta(3/2^+)$     & 0.635 & 0.51 & 0.390 & 0.537 \\
$\Delta^*(3/2^+)$   & 1.149 & 0.90 & 0.534 & $-$   \\
$\Omega(3/2^+)$     & 0.513 & $-$  & 0.395 & 0.418 \\
$\Omega^*(3/2^+)$   & 0.897 & $-$  & 0.543 & $-$   \\
\end{tabular}
\end{center}
\end{table}

\begin{table}[tbp]
\caption{Eigenvalue, in MeV, of the kinetic energy combined with different
contributions of the interacting potential. The subindexes in the 
potential stand for: $1=CON$, $2=CON+OGE$, $3=CON+\pi$,
$4=CON+OGE+\pi$, $5=CON+OGE+\pi +K$, $6=CON+OGE+\pi +K+\eta$,
$7=CON+OGE+\pi +K+\eta + \sigma$.}
\label{t2}
\begin{center}
\begin{tabular}{cccccccc}
State & $V_1$ & $V_2$ & $V_3$ & $V_4$ & $V_5$ & $V_6$ & $V_7$ \\ 
\hline
$N(1/2^+)$          &  1534  &  1254  &  1407  &  969  &   969  & 1030  &  939 \\  
$\Delta(3/2^+)$     &  1534  &  1314  &  1510  & 1291  &  1291  & 1283  & 1232 \\  
$N^*(1/2^+)$        &  1787  &  1601  &  1716  & 1448  &  1448  & 1479  & 1435 \\  
$N(1/2^-)$          &  1722  &  1530  &  1675  & 1422  &  1422  & 1447  & 1411 \\  
$\Sigma(1/2^+)$     &  1679  &  1417  &  1674  & 1408  &  1326  & 1229  & 1213 \\  
$\Sigma(3/2^+)$     &  1679  &  1462  &  1673  & 1454  &  1437  & 1438  & 1382 \\  
$\Sigma^*(1/2^+)$   &  1983  &  1757  &  1931  & 1752  &  1703  & 1688  & 1644 \\  
$\Sigma(1/2^-)$     &  1859  &  1677  &  1854  & 1671  &  1645  & 1634  & 1598 \\  
$\Lambda(1/2^+)$    &  1679  &  1405  &  1600  & 1225  &  1171  & 1217  & 1122 \\  
$\Xi(1/2^+)$        &  1819  &  1557  &  1819  & 1557  &  1472  & 1446  & 1351 \\
$\Omega(3/2^+)$     &  1955  &  1743  &  1955  & 1743  &  1743  & 1728  & 1650 \\  
\end{tabular}
\end{center}
\end{table}

\end{document}